\documentclass[
    ,final            % use final for the camera ready runs
  ]
  {aipproc}

\usepackage{amssymb,amsmath}

\layoutstyle{8x11single}

\newcommand{\beq}{\begin{equation}}
\newcommand{\eeq}{\end{equation}}

\begin{document}

\title{From Stopping to Viscosity in Nuclear Reactions}

\classification{21.65.-f, 25.75.-q, 25.75.Ag}
\keywords      {nuclear matter, shear viscosity, central reactions, transport theory, Boltzmann equation, stopping}

\author{Pawel Danielewicz}{
  address={
National Superconducting Cyclotron Laboratory and
Department of Physics and Astronomy\\ Michigan State University,
East Lansing, Michigan 48824, USA
}
}

\author{Brent Barker}{
  address={
National Superconducting Cyclotron Laboratory and
Department of Physics and Astronomy\\ Michigan State University,
East Lansing, Michigan 48824, USA
}
}

\author{Lijun Shi\footnote{Current address: National City Corporation, Cleveland, Ohio 44114, USA}}{
  address={
National Superconducting Cyclotron Laboratory and
Department of Physics and Astronomy\\ Michigan State University,
East Lansing, Michigan 48824, USA
}
}

\begin{abstract}
 Data on stopping in intermediate-energy central heavy-ion collisions are analyzed following transport theory based on the Boltzmann equation.  In consequence, values of nuclear shear viscosity are inferred.  The inferred values are significantly larger than obtained for free nucleon dispersion relations and free nucleon-nucleon cross sections.
\end{abstract}

\maketitle

%%%%%%%%%%%%%%%%%%%%%%%%%%%%%%%%%%%%%%%%%%%%
%% MAINMATTER
%%%%%%%%%%%%%%%%%%%%%%%%%%%%%%%%%%%%%%%%%%%%

\section{Introduction}

In central reactions of heavy nuclei, momentum is transferred between matter originating from opposing nuclei.  Towards the end of a reaction, the matter can be described as locally equilibrated, in terms of local temperature field folded with a field of local collective velocity.  Obviously, in a reaction, dissipation takes place.  Looking at appropriate data from the reactions, one can ask about the pace of dissipation and examine what that pace tells about the general dissipative properties of nuclear matter.

A theoretical model for the central reactions needs to be capable of describing different stages of a reaction, nonequilibrium and equilibrium.  Here, we shall rely on a single-particle description, in terms of the nucleon Wigner functions $f({\pmb p},{\pmb r},t)$ obeying a set of Boltzmann equations:
\beq
\label{eq:Boltzmann}
{\partial f_i \over \partial t} + {\partial \epsilon_{{\pmb p}_i} \over
\partial {\pmb p}_i }
\, {\partial f_i \over \partial {\pmb r}} -
{\partial \epsilon_{{\pmb p}_i} \over
\partial {\pmb r} } \, {\partial f_i \over \partial {\pmb p}_i}
= \sum_j
\int  { \text{d}{\pmb p}_j }
\, \text{d} \Omega' \,
v_{ij} \, \frac{\text{d} \sigma}{\text{d} \Omega'} \,
 \big( \tilde{f}_i \, \tilde{f}_j \,   f_i' \, f_j'
- \tilde{f}_i' \, \tilde{f}_j' \, f_i \, f_j \big) \, .
\eeq
In the above, the single-particle energy $\epsilon$ is a variational derivative \cite{Danielewicz:1999zn} of the energy represented as functional of the Wigner functions, $\epsilon = (2 \pi)^{-3} \, \delta E/ \delta f$.  The l.h.s.\ of \eqref{eq:Boltzmann} accounts for changes in the Wigner function due to the movement of particles, at velocity ${\pmb v} = \partial \epsilon/\partial {\pmb p}$, and due to their acceleration on account of the single-particle energy changing with position, with ${\pmb F} = - \partial \epsilon/\partial {\pmb r}$ representing a force on the particle.  The r.h.s.\ of the equation accounts for changes in $f$ due to collisions.  The pace of approach to local equilibrium is governed by cross-sections $\sigma$, most often assumed to coincide with those in free space.  The factors $\tilde{f}$ are Pauli-blocking factors, $\tilde{f} = 1 - f$.  In models of reactions where the Boltzmann set is not directly followed, it is still common to incorporate elements of the Boltzmann equation, such as the scattering governed by cross sections~\cite{Ono:2004}.

In comparing a transport model to reaction data, assumptions within the model are adjusted until a reasonable agreement reached.  Obviously, some data test some assumptions better and some worse.  An issue is the universality of drawn conclusions.  Information pertaining to a transient reaction stage alone or, even worse, just to specific model, can be of very limited utility.  For that reason, while a model is used to describe a reaction through its nonequilibrium stages, the same model is typically used to extrapolate the conclusions to those pertaining to finite- or zero-temperature equilibrium, and referring then to energy, pressure or optical potential \cite{Danielewicz:1999zn,Sturm:2000dm,Klahn:2006ir,Persram:2001dg,Danielewicz:2002pu}.  Here, see also~\cite{Gaitanos:2004ic,Danielewicz:2002he}, we shall consider characterization of the nuclear system for weak deviations from equilibrium, in terms of shear viscosity coefficient.  Just as the equation of state, the macroscopic transport coefficients might be assigned to nuclear matter even for models partly or fully phenomenological in nature, provided one could legitimately tie the variation in transport properties to particular observables.

The Boltzmann set \eqref{eq:Boltzmann} should apply to nuclear systems at low nucleon density $n$ combined with moderate to high temperatures~$T$.  At high densities, when the binary-collision regions would likely overlap, the equation provides just phenomenological description.  However, when the collisions are frequent in a system, the system approaches a local thermal equilibrium and begins to behave hydrodynamically.  In the latter limit, microscopic details behind the hydrodynamic behavior might not be important, as long as macroscopic properties are properly reproduced.

\section{Shear Viscosity}

\begin{figure}
\parbox[b]{.47\linewidth}{\includegraphics[width=\linewidth]{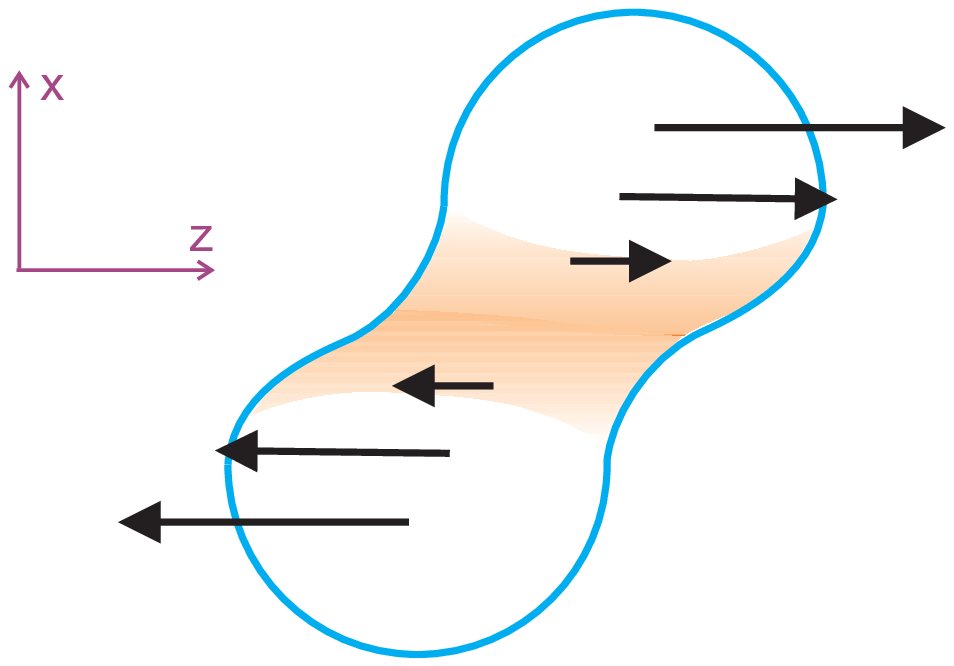}
}
\hspace*{.05in}
\parbox[b]{.23\linewidth}{
\includegraphics[width=\linewidth]{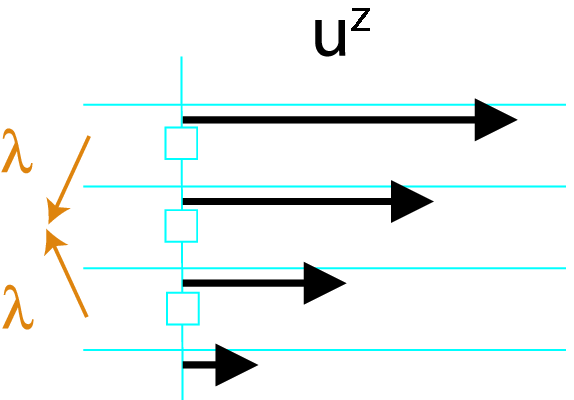}\\[3ex]
}
\hspace*{.3in}
\parbox[b]{.19\linewidth}{
\includegraphics[width=\linewidth]{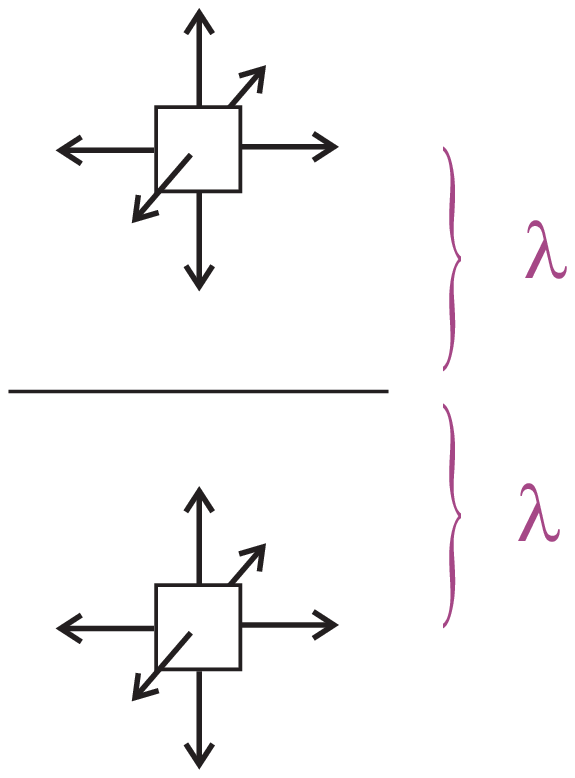}\\
}
\caption{
Schematic illustration for momentum transport in a reaction, between the projectile and target zones.  In a semiperipheral reaction, the collective velocity has predominantly longitudinal $z$-components and changes in transverse directions such as the $x$-direction along the reaction plane.  Associated with the velocity gradient, close to equilibrium, is the transport of the $z$-component of momentum in the $x$ direction.  Any location in the matter is reached by nucleons from about one mean-free-path $\lambda$ away, contributing to the momentum flux.
}
\label{fig:MomentumFlux}
\end{figure}

For weak gradients in a largely equilibrated system, fluxes of macroscopic quantities, leading to dissipation, are proportional to gradients within the system.  The Curie principle, stating that transformation properties for interrelated fluxes and gradients must be the same, allows for sorting out the possible relations.  The shear viscosity coefficient~$\eta$, that we shall be after, is the coefficient of proportionality between anisotropy of momentum-flux tensor, inducing dissipation, and velocity gradients.  When local velocity ${\pmb u}$ is directed along a specific $z$-direction, as is the case approximately in a semiperipheral nuclear reaction, see Fig.~\ref{fig:MomentumFlux}, and when ${\pmb u}$ changes in value in the perpendicular direction~$x$, the flux $\Pi^{zx}$, of $z$-component of momentum in $x$-direction, is
\beq
\Pi^{zx} = - \eta \, \frac{\partial u^z}{\partial x} \, ,
\label{eq:eta_definition}
\eeq
to linear order in gradients.

A simple estimate for the viscosity coefficient may be arrived at by carrying out mean-free-path considerations.  Thus, for transport in the $x$-direction, different locations along the $x$-axis will matter.  Any position along the $x$-axis, see Fig.~\ref{fig:MomentumFlux}, will be reached by particles that start out one mean free path $\lambda$ away.  Only about $1/6$ of particles at the starting location move, at typical speed~$v_\text{kin}$, towards the point of interest.  The average $z$-momentum, they bring in, is $m u^z$, where $u^z$ is that for their starting location.  Accounting for the fluxes due to particles moving up and down the $x$-axis yields
\beq
\Pi^{zx}   \simeq   \frac{1}{6} \, n \, v_\text{kin} \, m \, u^z(x -
\lambda) - \frac{1}{6} \, n \, v_\text{kin} \, m \, u^z(x + \lambda)
  \simeq - \, \frac{1}{3} \, n \, v_\text{kin} \, m \, \lambda \,
\frac{\partial u^z}{\partial x} \, ,
\eeq
where $n$ is particle density.  This yields then an estimate for the viscosity coefficient of
\beq
\eta  \simeq   \frac {n \, m \, v_\text{kin}}{3} \, \lambda
 \sim  \frac{0.16 \, \mbox{fm}^{-3} \, 939 \, \mbox{MeV}/c^2 \, 0.3 \, c}{3}
\, 2\, \mbox{fm} \sim 30 \, \mbox{MeV/fm}^2c \, ,
\eeq
where, in the quantitative estimate, we have used values representative for reactions at moderate beam energies.  Proportionality of the viscosity to the mean free path implies that the viscosity is inversely proportional to the interaction cross sections, or otherwise exhibits a negative correlation with stopping, since $\lambda \sim 1/(n \sigma)$.

When comparing data to the results of the Boltzmann set, to infer nuclear viscosity, one needs to determine first the viscosity for the set.  For this, one must consider a system governed by \eqref{eq:Boltzmann}, close to equilibrium.  The r.h.s.\ of the equations in the set vanishes for local equilibrium distributions,
\beq
f_i^\text{eq} = \frac{1}{\mbox{exp}\left(\frac{ \frac{({\pmb p} - m \, {\pmb u})^2}{2m} -\mu_i}{T}\right)+1} \, ,
\label{eq:f_eq}
\eeq
where ${\pmb u}$, $\mu_i$ and $T$ are equilibrium parameters that depend on position.  However, in the latter case, the distributions cannot solve the set, because of the derivatives evaluated on the l.h.s.  With \eqref{eq:Boltzmann}, the local equilibrium distributions~\eqref{eq:f_eq} need to be to corrected to at least the first order in gradients in equilibrium parameters, for these distributions to solve the Boltzamnn set.  Notably, the local-equilibrium distributions \eqref{eq:f_eq} would yield vanishing dissipative contributions to macroscopic fluxes, in particular to the anisotropy of the tensor of momentum flux.  On the other hand, corrections linear in gradients will yield contributions to fluxes that are linear in gradients, that are of interest in the context of \eqref{eq:eta_definition}.  Following the Chapman-Enskog method, one can, in fact, systematically seek a solution to \eqref{eq:Boltzmann} around local equilibrium, by expanding the distributions in derivatives
\beq
f_i = f_i^{(0)} + f_i^{(1)} +  f_i^{(2)} + \ldots \,
\eeq
where $ f_i^{(0)} \equiv f^{eq}$ and $f^{(n)}$ is of the $n$'th order in gradients.  The $n$'th order corrections may be obtained from the Boltzmann set by inserting the $(n-1)$'th terms into the l.h.s.

For inferring shear viscosity, the $f_i^{(1)}$ terms are important.  Upon inserting $f_i^\text{eq}$ to the l.h.s.\ of \eqref{eq:Boltzmann}, the form of $f_i^{(1)}$ emerges \cite{Shi:2003np,Danielewicz:1984kt}:
\beq
f_i^{(1)} = \phi_i \, f_i^{(0)} \, (1 - {f}_i^{(0)})
\eeq
where $\phi_i$ are smooth functions of position and momenta.  At low temperatures, the distributions are modified, compared to local equilibrium, in the region of Fermi surface.  Following the Curie principle, anisotropy of symmetric momentum-flux tensor should be driven by the anisotropy of symmetric tensor of velocity gradient:
\beq
\phi_i  =  b_i \, \left( p_k \, p_\ell - \frac{p^2}{3} \, \delta_{k \ell} \right)
\,
\left(\nabla_k \, u_{\ell} + \nabla_\ell \, u_k - \frac{2}{3} \, \delta_{k \ell} \,
\nabla_n \, u_n \right) \, ,
\eeq
where $b_i$ are, generally, functions of momentum magnitude.  Assuming that $b_i$ change weakly momentum, one can arrive at a closed expression \cite{Shi:2003np,Danielewicz:1984kt} for the viscosity, accurate in practice down to few percent for a given cross section, and provided here in the simplified form for symmetric matter:
\beq
\eta= \frac{5T}{9} \frac{\left( \int \text{d}{\bf p} \, f \, p^2 \right)^2}
{\int \text{d}{\bf p}_1 \, \text{d}{\bf p}_2 \, \text{d}\Omega \,
f_1 \, f_2 \, (1 - f_1') \, (1 - f_2') \, v_{12} \,
\frac{\text{d} \sigma}{\text{d} \Omega} \, p_{12}^4 \, \sin^2{\theta}} \, .
\label{eq:eta=}
\eeq

\begin{figure}
{\includegraphics[width=.78\linewidth]{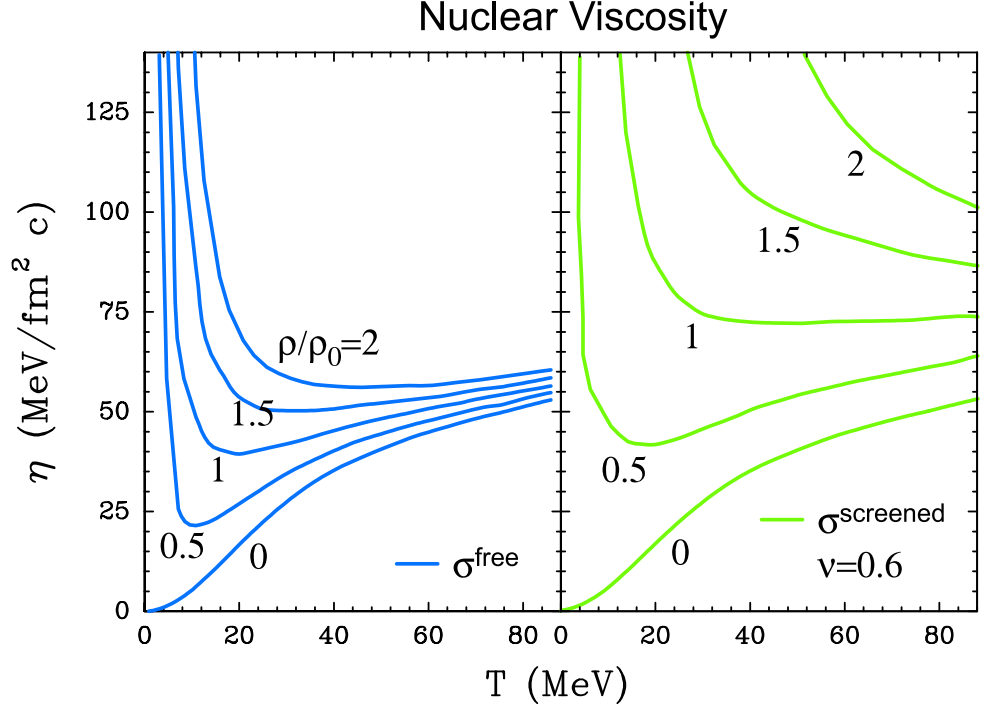}
}
\caption{Viscosity in symmetric nuclear matter at different densities, as a function of temperature.  Left panel shows viscosity calculated with free-space cross-sections and velocities.  Right panel shows viscosity calculated with in-medium cross-sections and effective masses adjusted to heavy-ion data.
}
\label{fig:Viscosity}
\end{figure}

The shear viscosity coefficient indeed comes out in \eqref{eq:eta=} inversely proportional to the cross sections.  However, the cross-sections in the viscosity get weighted with Pauli-blocking factors.  Additional weighting there indicates that the cross sections at high relative-momenta and large scattering-angles matter more for the viscosity than the cross sections at low momenta and small angles.  The left panel in Fig.~\ref{fig:Viscosity} shows the shear viscosity from Boltzmann equation, calculated with cross sections and velocities such as in free space.  For typical conditions in a reaction, at moderate incident energies, the calculated viscosities are higher than in the simple mean-free-path estimate.  At low temperatures, the viscosity values diverge due to Pauli blocking of collisions.

We shall next examine whether data pertinent to dissipation of momentum \cite{Reisdorf:2004wg,PhysRevC.57.R1032} justify the use of free elementary cross sections in the description of nuclear dynamics.  Other types of data on nuclear dissipation, of general interest especially in the context of exotic beams, are those on equilibration of neutron-proton asymmetry \cite{Shi:2003np,PhysRevLett.92.062701,Rami:1999xq,Baran:2005ce}.  Previous data analyses have established a lowering of nucleon effective masses with increase in nuclear density \cite{Danielewicz:1999zn,Persram:2001dg}.

\section{Data Comparisons}

\subsection{$vartl$ Observable}

The first data, that we are going to compare our theory to, are those on the degree of isotropization of momentum distributions in symmetric central collisions.  To quantify the degree of isotropy in the final-state of a reaction, the FOPI Collaboration \cite{Reisdorf:2004wg} has introduced the observable
\beq
vartl = \frac{\Delta \, y_t}{\Delta \, y_l} \, ,
\eeq
that is the ratio of the widths of rapidity distributions in the transverse and longitudinal directions.  The rapidity for the longitudinal direction is defined in the standard manner.  The rapidity for the transverse direction is defined by replacing the beam direction, in the standard definition, by a random transverse direction.  Reaching isotropy in a reaction would produce $vartl \simeq 1$.  Transparency effects would yield $vartl < 1$.  Finally, a strong hydrodynamic behavior might yield $vartl > 1$ in a central reaction.  As systems should evolve, from transparency towards a hydrodynamics behavior, with increasing system mass, $vartl$ is expected to grow with system mass.

\begin{figure}
{\includegraphics[width=.49\linewidth,height=.32\textheight]{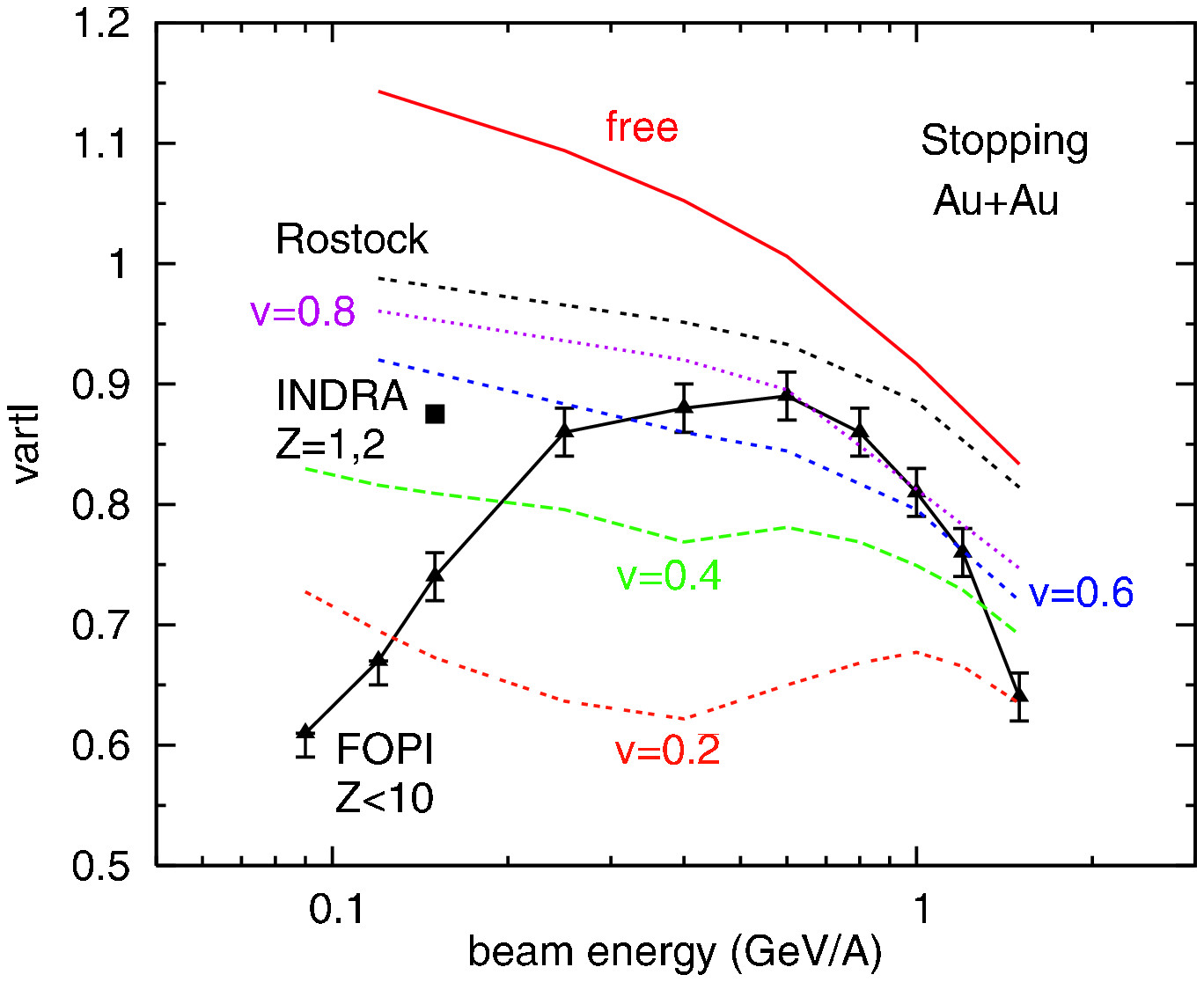}
}
\hspace*{.8em}
{\includegraphics[width=.45\linewidth,height=.32\textheight]{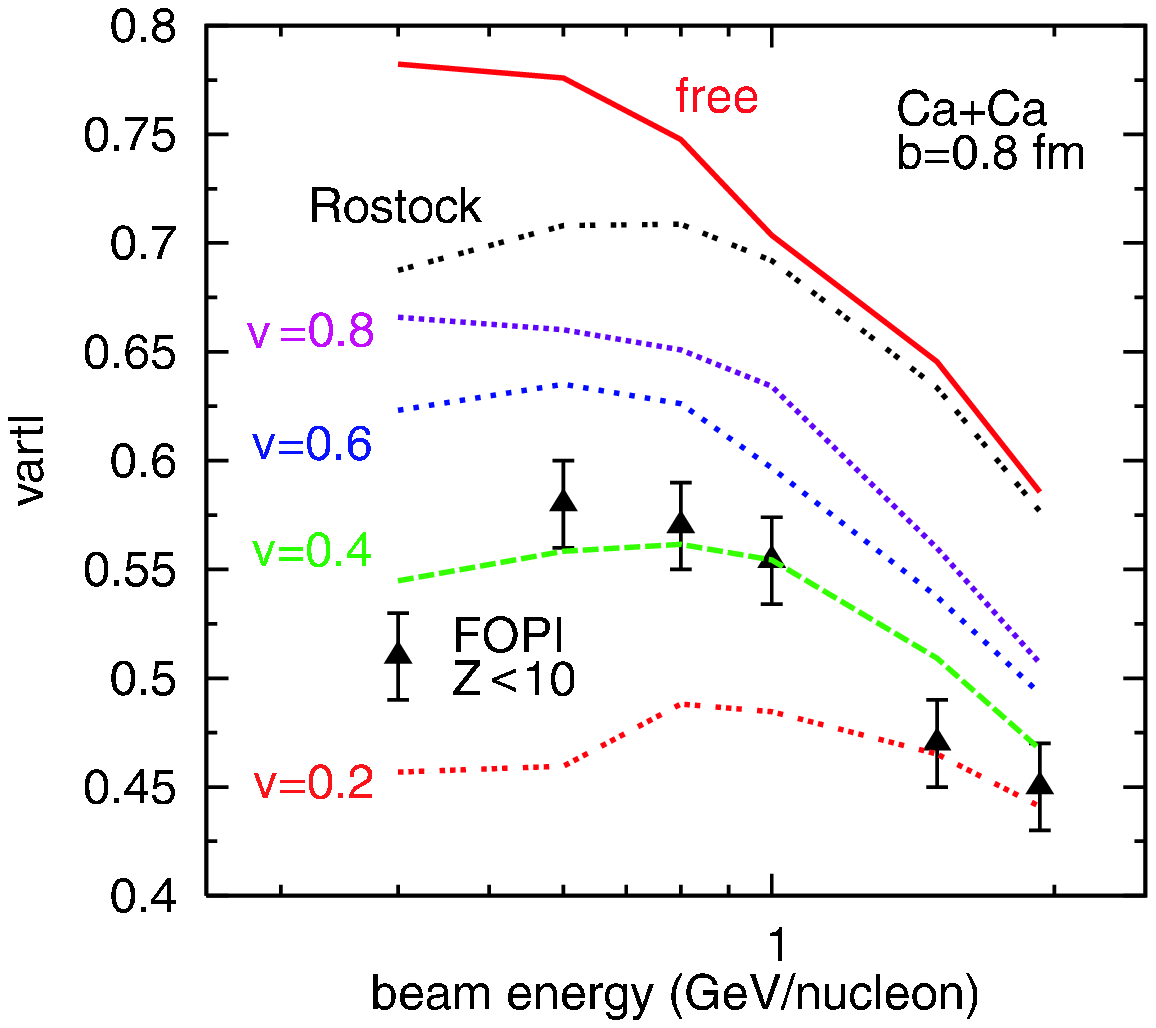}
}
\caption{Excitation function of the $vartl$ observable in central Au + Au (left panel) and Ca + Ca (right panel) collisions.  Symbols represent data of the FOPI \cite{Reisdorf:2004wg} (triangles) and INDRA \cite{Andronic:2006ra} (square) Collaborations.  The FOPI results are obtained using all fragments with charge number $Z < 10$.  The INDRA result is for $Z = 1$ and 2.  Lines represent results from the Boltzmann equation model including fragments with mass number $A \le 3$, for different assumptions on elastic elementary cross sections.
}
\label{fig:vartl}
\end{figure}

Excitation functions for central Au + Au and Ca + Ca collisions, from measurements of the FOPI Collaboration~\cite{Reisdorf:2004wg}, are shown in Fig.~\ref{fig:vartl}.  The FOPI Collaboration has calculated the $vartl$ values using all fragments with $Z < 10$.  To illustrate the effect of including the fragments with relatively high-$Z$ values, we include also an Au + Au result from the
INDRA Collaboration \cite{Andronic:2006ra}, obtained including fragments with $Z = 1$ and 2 only.  In central collisions, the intermediate-mass-fragments with $Z \ge 3$ are abundant at beam energies around 100~MeV/nucleon, but become less frequent at 400~MeV/nucleon and above.  Also, such fragments are more frequent in a heavy system such as Au + Au than the relatively light Ca + Ca.

As is apparent in Fig.~\ref{fig:vartl}, when looking at the $Z < 10$ fragments, the proximity to isotropy is observed only at intermediate energies in Au + Au collisions, with $vartl \gtrsim 0.85$ there at $400 - 800 \, \text{MeV/nucleon}$.  At low energies, the original nuclei might not interpenetrate enough for the momentum dissipation to get completed and Pauli principle might further suppress collisions.  At high energies, elementary cross sections become forward peaked, leading to transparency.

To facilitate comparisons to data, we include in the model calculations, at low densities, the production of deuterons, tritons and helions in few-nucleon collisions \cite{Danielewicz:1992mi}.  Since, however, the model does not predict yields of heavier fragments, the comparisons to the FOPI data will not be meaningful below 250~MeV/nucleon for Au + Au reactions.  Also, in comparisons, we will manipulate only elastic elementary cross sections only, dominating reactions below 800~MeV/nucleons.  Thus, most relevant conclusions will pertain to the energy region of $250 - 800 \, \text{MeV/nucleon}$.

It is apparent in the left panel of Fig.~\ref{fig:vartl}, that the model calculations with free cross sections grossly overestimate stopping data, yielding even values of $vartl > 1$ below 800~MeV/nucleon, in the indication of a strong hydrodynamic behavior.  Clearly, the free cross sections produce excessive stopping in those reactions.  Subsequent issue is of how the cross sections should be modified and, more accurately, reduced in the medium.  There had been calculations in the literature, carried out for equilibrated nuclear-matter, of changes in the cross sections due to effects of effective mass and effects of the Pauli principle on intermediate states in two-nucleon scattering \cite{PhysRevC.55.3006,PhysRevC.57.806,PhysRevC.64.024003}.  General finding has been that of lowering of the low-energy nucleon-nucleon cross sections in the medium.  Testing those cross sections, we carry out Boltzmann-equation simulations employing parameterized results on the in-medium reduction of the cross-sections, by the Rostock group and their collaborators  \cite{PhysRevC.55.3006,PhysRevC.57.806}.  The $vartl$ results shown in Fig.~\ref{fig:vartl} are reduced compared to those for the free cross-sections, but they are still excessive compared to the data.

The nuclear-matter calculations of cross sections, such as \cite{PhysRevC.55.3006,PhysRevC.57.806,PhysRevC.64.024003}, do not account for overlapping of the  collision regions at high densities, that compete against each other.  To account phenomenologically for a~unitary saturation taking place, we require that the nucleon-nucleon cross section cannot exceed a size imposed by interparticle distances
\beq
\sigma \lesssim \sigma_0 = \nu \, n^{-2/3} \, ,
\label{eq:sigmanu}
\eeq
where $\nu$ is a parameter of the order of~1.  To realize this limit in practice, we parameterize the in-medium cross with
\beq
\sigma = \sigma_0 \, \tanh{\left(\frac{\sigma_\text{free}}{\sigma_0}\right)} \, ,
\label{eq:sigma0}
\eeq
where $\nu$ is adjusted.  In the low-density limit, $n \rightarrow 0$, the in-medium cross-section approaches then the free cross-section, $\sigma \rightarrow \sigma_\text{free}$.  At high-density, $n \rightarrow \infty$, the in-medium cross section approaches $\sigma_0$ from below, $\sigma \nearrow \sigma_0$.  From the Au + Au results for different $\nu$ in Fig.~\ref{fig:vartl}, it is apparent that $\nu = 0.6$ is closest to the data in the relevant energy region.

If we next turn to the Ca + Ca $vartl$ results in Fig.~\ref{fig:vartl}, we find again that calculations with free cross sections again strongly overestimate the measurements.  However, the $\nu=0.6$ calculations, favored in the Au + Au case, also overstimate the data.  The Ca + Ca data rather favor $\nu = 0.4$.  However, an issue in the Ca + Ca reactions is of a less precise determination of the centrality of reactions than in Au + Au.  Lower multiplicities in Ca + Ca collisions, than in Au + Au, produce fluctuations for observables used to constrain reaction centrality.  If we assume that the collisions in the measurements have not been fully central, we can get agreement between calculations and data for $\nu \simeq 0.6$.

\subsection{Linear Momentum Transfer}

\begin{figure}
{\includegraphics[width=.77\linewidth]{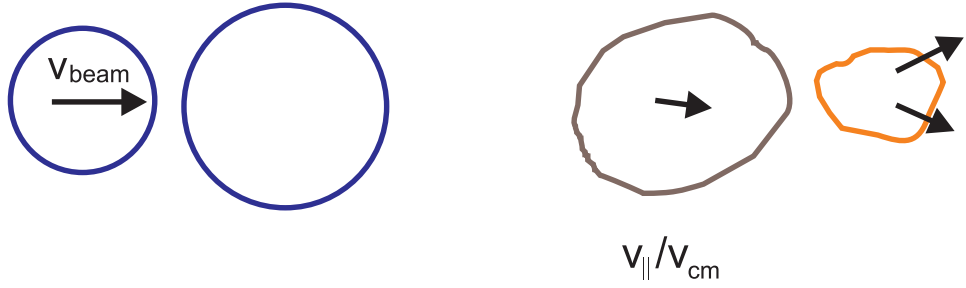}
}
\caption{Schematic illustration of measurements of linear momentum transfer.  Among products of the reaction, the most massive fragment is identified.  Its longitudinal velocity is compared to the center of mass velocity for the system.
}
\label{fig:LMT}
\end{figure}

Another observable, pertinent to stopping, has been the linear momentum transfer (LMT), used, in particular, for quantifying central asymmetric collisions of heavy ions.  In the measurements~\cite{PhysRevC.57.R1032}, the heaviest emitted fragment has been identified, see Fig.~\ref{fig:LMT}.  Most likely that fragment originates from the target residue left after the fast initial stage of the reaction.  The average longitudinal component of the fragment velocity should reflect the average velocity of the residue.  A scale-invariant measure of the degree of stopping in a reaction may be obtained by taking a ratio of the average fragment velocity to the center-of-mass velocity for the system as a whole.  If the target and projectile fuse, then the ratio should be close to~1, $\langle v_\parallel \rangle /v_\text{cm} \simeq 1$.  Little stopping should be characterized by $\langle v_\parallel \rangle /v_\text{cm} \ll 1$.

\begin{figure}
{\includegraphics[width=.48\linewidth,height=.28\textheight]{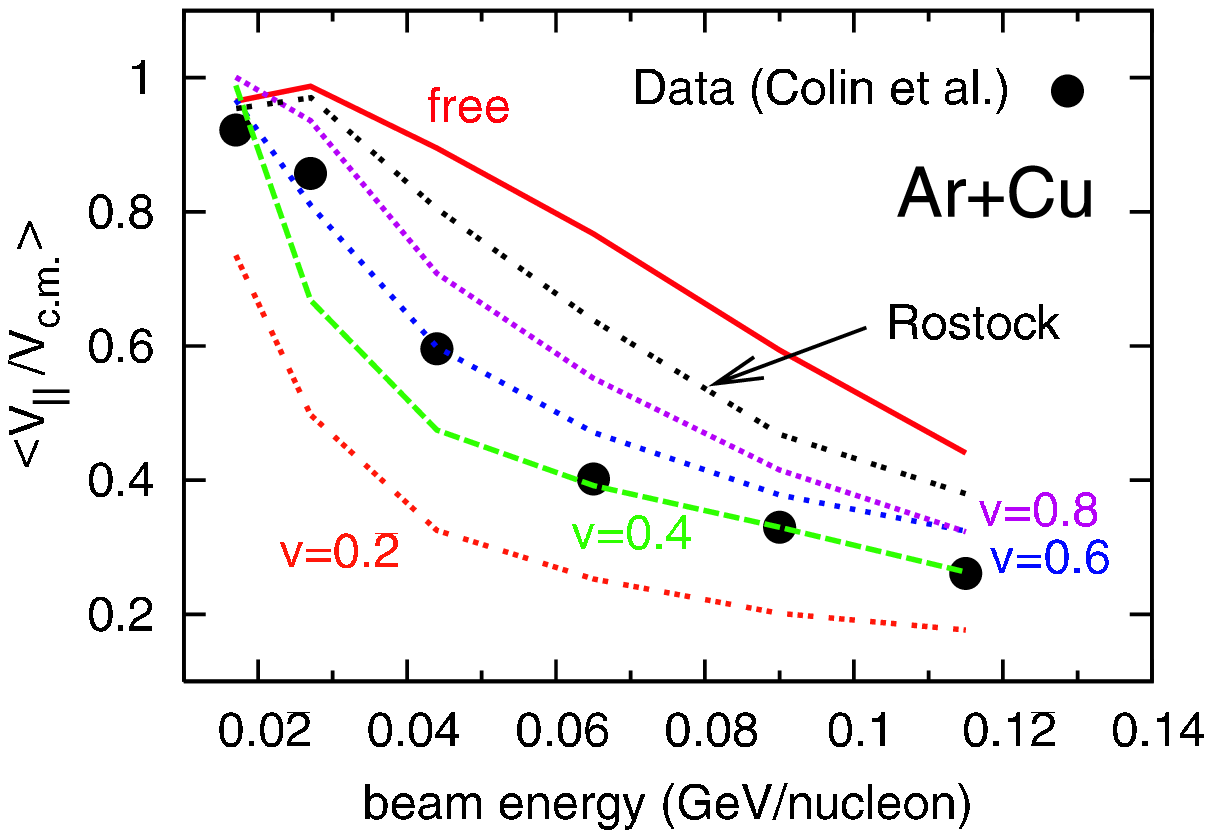}
}
\hspace*{.2em}
{\includegraphics[width=.48\linewidth,height=.28\textheight]{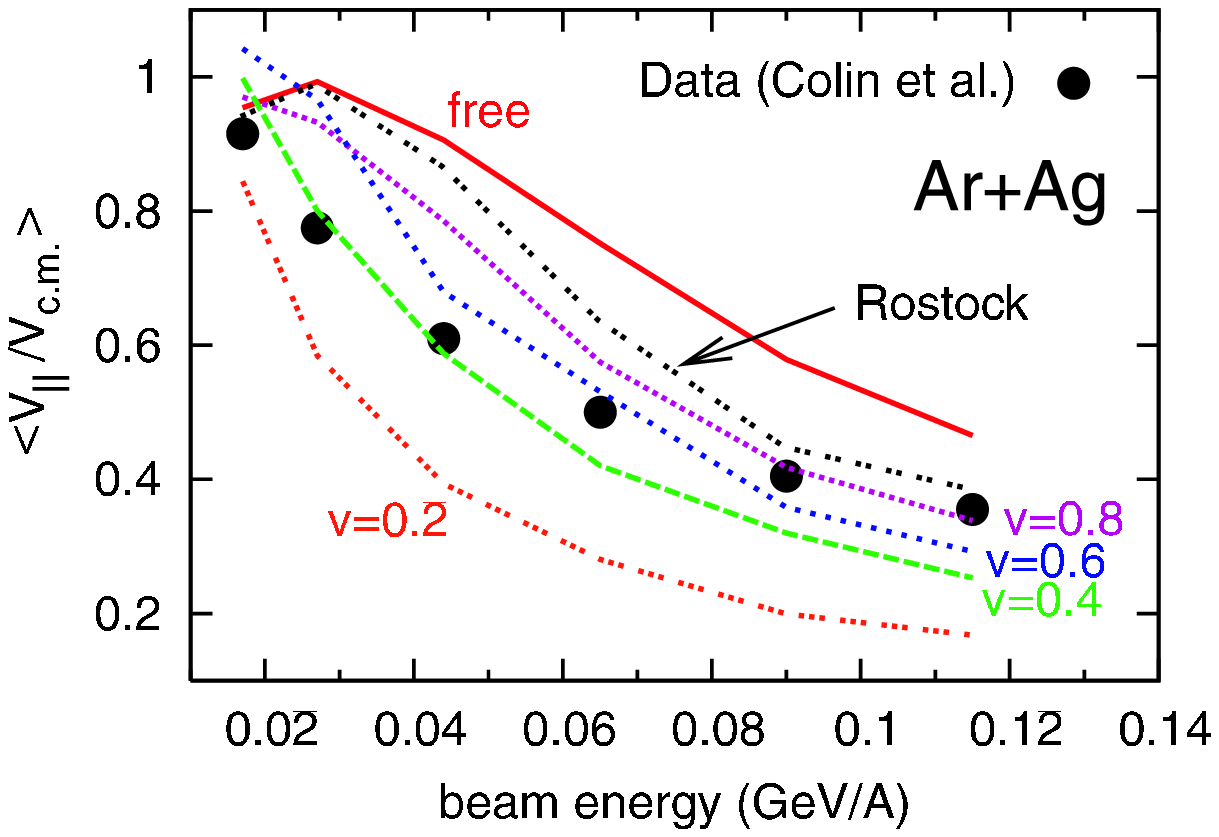}
}
\caption{Excitation energy for the measure of stopping $\langle v_\parallel/v_\text{cm} \rangle$ in central Ar + Cu (left panel) and Ar + Ag (right panel) collisions.  The velocity component $v_\parallel$ is that of the heaviest fragment emitted from the reaction.  Symbols represent data of Ref.~\cite{PhysRevC.57.R1032} and lines represent results of Boltzmann-equation simulations employing different assumptions on elementary in-medium cross sections.
}
\label{fig:limot}
\end{figure}

Figure \ref{fig:limot} shows measured and calculated excitation functions for $\langle v_\parallel \rangle /v_\text{cm}$ in central Ar + Ag and Ar + Cu collisions.  At low beam energies, the nuclei appear to fuse.  As beam energy increases stronger and stronger transparency sets in.  Calculations with free cross definitely overestimate the stopping in the reactions.  The Rostock in-medium cross sections reduce stopping, but not quite enough.  For the cross-sections of Eqs.~\eqref{eq:sigma0} and \eqref{eq:sigmanu}, getting close to the data requires going down in $\nu$ to the vicinity of $\nu = 0.6$.

\section{Nature of Conclusions}

\begin{figure}
{\includegraphics[width=.47\linewidth,height=.28\textheight]{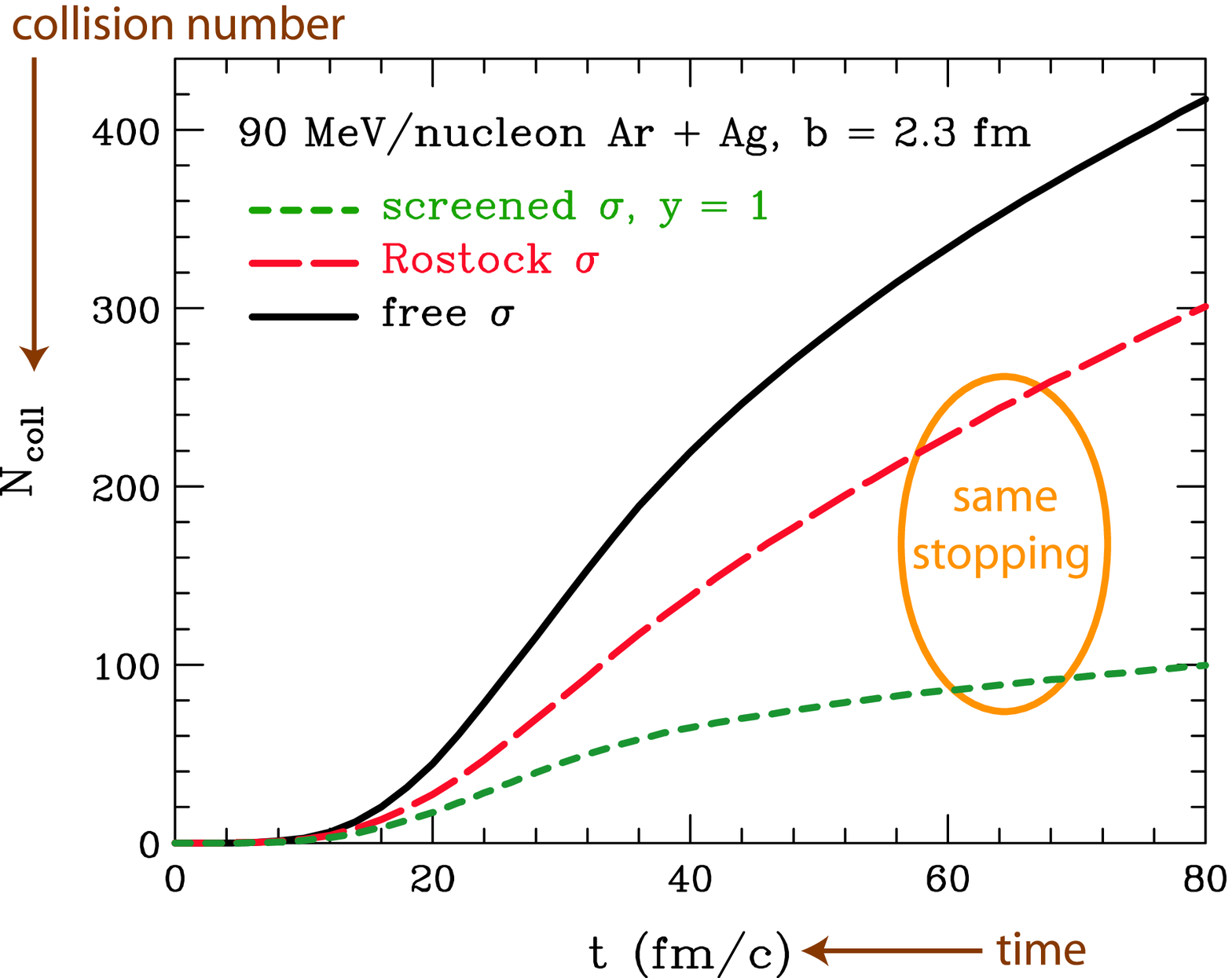}
}
\hspace*{.5em}
{\includegraphics[width=.47\linewidth,height=.28\textheight]{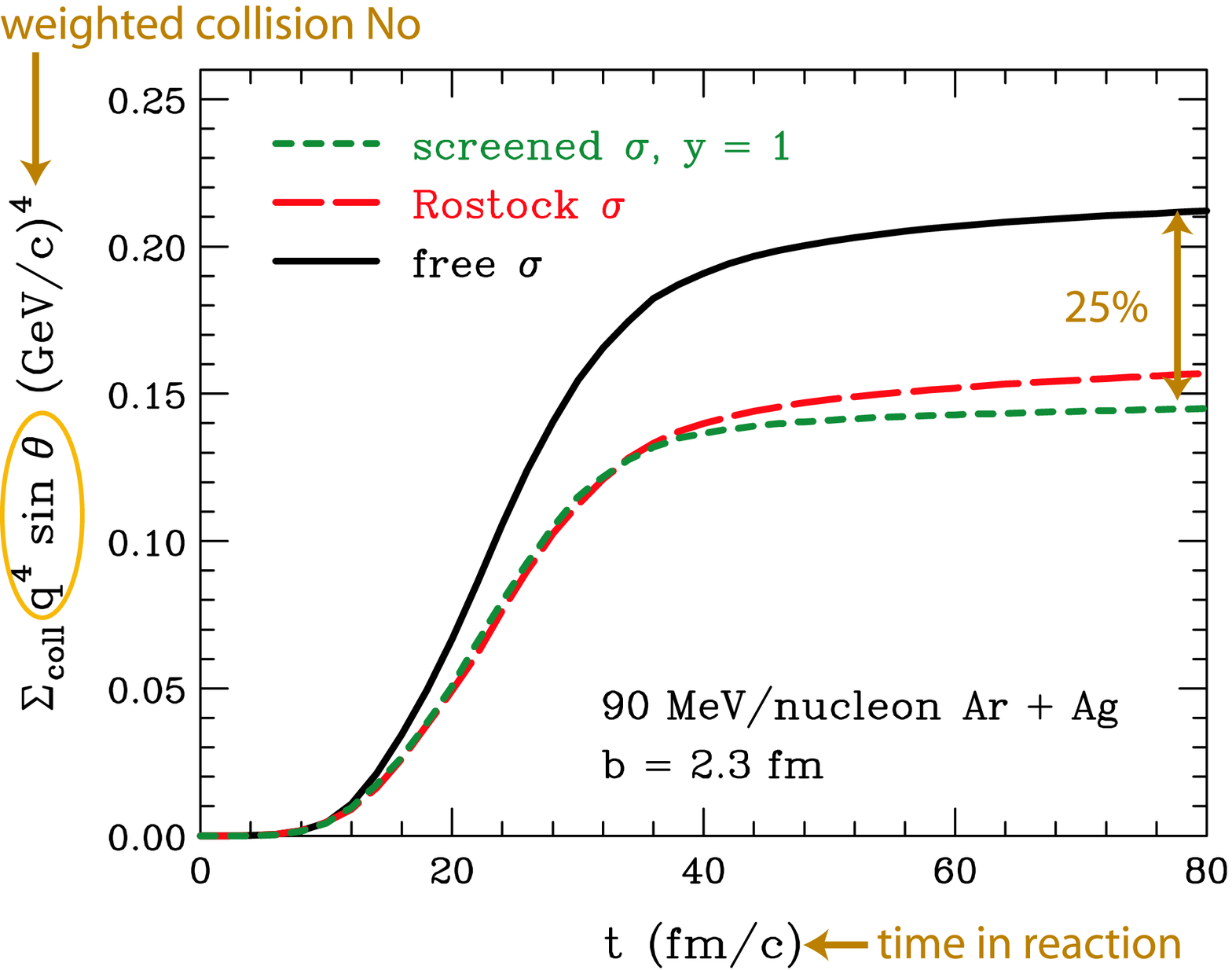}
}
\caption{Measures of elementary collisions in the simulations of central Ar + Ag reaction at 90~MeV/nucleon, for different assumptions on elementary cross sections, as a function of time~$t$ in the reaction.  The left panel shows net collision number.  The right panel shows the number of collisions weighted with the weight $p_{12}^4 \sin{\theta}^2$ that is applied to the collision rate in the expression~\eqref{eq:eta=} for viscosity.
}
\label{fig:colli}
\end{figure}

It is apparent that the use of free cross sections in the Boltzmann-equation simulations yields too much stopping.  The nature of the implied in-medium reduction of cross-sections, however, is not obvious. Specifically, different types of cross-section reduction can produce similar stopping.  Thus, with $\nu \simeq 1$ in Eqs.~\eqref{eq:sigmanu} and \eqref{eq:sigma0}, we can get nearly identical stopping results in terms of $vartl$ and $\langle v_\parallel/v_\text{cm} \rangle$, for the reactions in Figs.~\ref{fig:vartl} and \ref{fig:limot}.  However, collision counts for the simulations utilizing the two cross-section reductions are then vastly different, see the left panel in Fig.~\ref{fig:colli} for results for the Ar + Ag reaction at 90~MeV/nucleon.  Compared to the simulation with free cross sections, the collision number drops just by 25\% and by a factor of~4, respectively, for the Rostock and $\nu = 1$ cross sections, respectively!

As we have stated, however, it is not clear to what extent the separate nucleon-nucleon collisions can be, in reality, identified within a reaction.  Even if they can be identified, some, especially those at low relative momenta and those with forward scattering angles, may matter little for the reaction dynamics.  Around equilibrium, solely macroscopic properties of matter would be important for the dynamics and the same dynamics would result from different cross sections as long as those cross sections gave the same transport coefficients.  For dissipation of momentum, the dominant role would be played by the viscosity coefficient and, in fact, that coefficient might not even be tied to a medium for which the kinetic limit applies.  In the coefficient for the kinetic limit, Eq.~\eqref{eq:eta=}, the cross section is multiplied by the weight $p_{12}^4 \sin{\theta}^2$, which emphasizes collisions at high relative momentum, populating wide angles.  Collisions weighted in this fashion should matter for dissipation of momentum close to equilibrium.  The right panel in Fig.~\ref{fig:colli} next shows the collision numbers for different cross sections weighted with the viscous weight.  The weighted collision numbers are now similar for the two in-medium cross-sections that produce similar stopping, consistently with the discussion above.  With a similar count of the weighted collisions, the two cross would produce similar shear viscosity coefficients for conditions such as in the reactions.

The right panel in Fig.~\ref{fig:Viscosity} shows the shear viscosity coefficient calculated with the $\nu=0.6$ cross-section values, that yield stopping of such order as observed, and with in-medium dispersion relations established earlier \cite{Danielewicz:1999zn}.  Both the lowering of cross sections and of effective masses contribute to an enhancement of the coefficient values, compared to the results obtained when disregarding in-medium effects, in the left panel of Fig.~\ref{fig:Viscosity}.

\section{Summary}

We have confronted the measurements of stopping in central nuclear collisions, in terms of the $vartl$ observable and the linear-momentum transfer, with the predictions of stopping from a~Boltzmann-equation set.  Predictions utilizing free nucleon-nucleon cross-sections strongly overestimate the stopping, irrespectively of the considered observable or the system.  Predictions utilizing cross-sections from nuclear-matter calculations \cite{PhysRevC.55.3006,PhysRevC.57.806,PhysRevC.64.024003} yield less stopping than for the free cross sections, but still too much compared to data.  The stopping data alone do not constrain details in the in-medium cross-section alone.  Similar stopping predictions can be arrived at within calculations where collision numbers differ by a factor of~3.  Fortunately, for drawing conclusions from the stopping data, similar stopping results are correlated to similar predictions for nuclear shear viscosity.  With inclusion of the in-medium effects, the deduced viscosity values are significantly larger than anticipated in the absence of such effects.  Systematics of the nuclear viscosity values is, in particular, of interest in assessing how a nuclear system may approach the limit of near-perfect liquid with increase of temperature \cite{Lacey:2006bc}.  The consideration of the latter limit involves also analysis of entropy values that may be deduced from fragment yields in the reactions.

\begin{theacknowledgments}
This work was supported by the National Science Foundation under Grants PHY-0555893 and PHY-0800026.
\end{theacknowledgments}

\bibliographystyle{aipproc}   % if natbib is available

\bibliography{coel08}

\begin{thebibliography}{21}
\expandafter\ifx\csname natexlab\endcsname\relax\def\natexlab#1{#1}\fi
\providecommand{\enquote}[1]{``#1''}
\expandafter\ifx\csname url\endcsname\relax
  \def\url#1{\texttt{#1}}\fi
\expandafter\ifx\csname urlprefix\endcsname\relax\def\urlprefix{URL }\fi
\providecommand{\eprint}[2][]{\url{#2}}

\bibitem[Danielewicz(2000)]{Danielewicz:1999zn}
P.~Danielewicz, \emph{Nucl. Phys.} \textbf{A673}, 375--410 (2000),
  \eprint{nucl-th/9912027}.

\bibitem[Ono and Horiuchi({2004})]{Ono:2004}
A.~Ono, and H.~Horiuchi, \emph{{Prog. Part. Nucl. Phys.}} \textbf{{53}},
  {501--581} ({2004}).

\bibitem[Sturm et~al.(2001)]{Sturm:2000dm}
C.~T. Sturm, et~al., \emph{Phys. Rev. Lett.} \textbf{86}, 39--42 (2001),
  \eprint{nucl-ex/0011001}.

\bibitem[Klahn et~al.(2006)]{Klahn:2006ir}
T.~Klahn, et~al., \emph{Phys. Rev.} \textbf{C74}, 035802 (2006),
  \eprint{nucl-th/0602038}.

\bibitem[Persram and Gale(2002)]{Persram:2001dg}
D.~Persram, and C.~Gale, \emph{Phys. Rev.} \textbf{C65}, 064611 (2002),
  \eprint{nucl-th/0111035}.

\bibitem[Danielewicz et~al.(2002)]{Danielewicz:2002pu}
P.~Danielewicz, R.~Lacey, and W.~G. Lynch, \emph{Science} \textbf{298},
  1592--1596 (2002), \eprint{nucl-th/0208016}.

\bibitem[Gaitanos et~al.(2005)]{Gaitanos:2004ic}
T.~Gaitanos, C.~Fuchs, and H.~H. Wolter, \emph{Phys. Lett.} \textbf{B609},
  241--246 (2005), \eprint{nucl-th/0412055}.

\bibitem[Danielewicz(2002)]{Danielewicz:2002he}
P.~Danielewicz, \emph{Acta Phys. Polon.} \textbf{B33}, 45--64 (2002),
  \eprint{nucl-th/0201032}.

\bibitem[Shi and Danielewicz(2003)]{Shi:2003np}
L.~Shi, and P.~Danielewicz, \emph{Phys. Rev.} \textbf{C68}, 064604 (2003),
  \eprint{nucl-th/0304030}.

\bibitem[Danielewicz(1984)]{Danielewicz:1984kt}
P.~Danielewicz, \emph{Phys. Lett.} \textbf{B146}, 168--175 (1984).

\bibitem[Reisdorf et~al.(2004)]{Reisdorf:2004wg}
W.~Reisdorf, et~al., \emph{Phys. Rev. Lett.} \textbf{92}, 232301 (2004),
  \eprint{nucl-ex/0404037}.

\bibitem[Colin et~al.(1998)]{PhysRevC.57.R1032}
E.~Colin, et~al., \emph{Phys. Rev. C} \textbf{57}, R1032--R1036 (1998).

\bibitem[Tsang et~al.(2004)]{PhysRevLett.92.062701}
M.~B. Tsang, et~al., \emph{Phys. Rev. Lett.} \textbf{92}, 062701 (2004).

\bibitem[Rami et~al.(2000)]{Rami:1999xq}
F.~Rami, et~al., \emph{Phys. Rev. Lett.} \textbf{84}, 1120--1123 (2000),
  \eprint{nucl-ex/9908014}.

\bibitem[Baran et~al.(2005)]{Baran:2005ce}
V.~Baran, M.~Colonna, M.~Di~Toro, M.~Zielinska-Pfabe, and H.~H. Wolter,
  \emph{Phys. Rev.} \textbf{C72}, 064620 (2005), \eprint{nucl-th/0506078}.

\bibitem[Andronic et~al.(2006)]{Andronic:2006ra}
A.~Andronic, J.~Lukasik, W.~Reisdorf, and W.~Trautmann, \emph{Eur. Phys. J.}
  \textbf{A30}, 31--46 (2006), \eprint{nucl-ex/0608015}.

\bibitem[Danielewicz and Pan(1992)]{Danielewicz:1992mi}
P.~Danielewicz, and Q.-B. Pan, \emph{Phys. Rev.} \textbf{C46}, 2002--2011
  (1992).

\bibitem[Schulze et~al.(1997)]{PhysRevC.55.3006}
H.-J. Schulze, A.~Schnell, G.~R\"opke, and U.~Lombardo, \emph{Phys. Rev. C}
  \textbf{55}, 3006--3014 (1997).

\bibitem[Schnell et~al.(1998)]{PhysRevC.57.806}
A.~Schnell, G.~R\"opke, U.~Lombardo, and H.-J. Schulze, \emph{Phys. Rev. C}
  \textbf{57}, 806--810 (1998).

\bibitem[Fuchs et~al.(2001)]{PhysRevC.64.024003}
C.~Fuchs, A.~Faessler, and M.~El-Shabshiry, \emph{Phys. Rev. C} \textbf{64},
  024003 (2001).

\bibitem[Lacey et~al.(2007)]{Lacey:2006bc}
R.~A. Lacey, et~al., \emph{Phys. Rev. Lett.} \textbf{98}, 092301 (2007),
  \eprint{nucl-ex/0609025}.

\end{thebibliography}

\end{document}